\documentclass[preprint,aps,onecolumn]{revtex4-1}
\usepackage{amsmath,color}

\begin{document}

\title{Vacuum energy in Kerr-AdS black holes}
\author{\textbf{Gonzalo Olavarria}}
\email{gonzalo.olavarria.u@mail.pucv.cl}
\affiliation{ Instituto de F\'{i}sica, Pontificia Universidad Cat\'{o}lica de
Valpara\'{i}so, \\ Casilla 4059, Valpara\'{i}so, Chile, }

\author{\textbf{Rodrigo Olea}}
\email{rodrigo.olea@unab.cl}
\affiliation{Departamento de Ciencias F\'{i}sicas, Universidad Andres Bello, \\
Rep\'{u}blica 220, Santiago, Chile.}

\date{\today}

\begin{abstract}
We compute the vacuum energy for Kerr black holes with anti-de Sitter (AdS)
asymptotics in dimensions $5\leq D\leq 9$ with all rotation parameters. The
calculations are carried out employing an alternative regularization scheme
for asymptotically AdS gravity, which considers supplementing the bulk
action with counterterms which are a given polynomial in the extrinsic and
intrinsic curvatures of the boundary (also known as Kounterterms). The
Kerr-Schild form of the rotating solutions enables us to identify
the vacuum energy as coming from the part of the metric that corresponds to a
global AdS spacetime written in oblate spheroidal coordinates. We find
that the zero-point energy for higher-dimensional Kerr-AdS reduces to one of a Schwarzschild-AdS black hole when all the rotation parameters are
equal to each other, a fact that is well known in five dimensions. We also
sketch a compact expression for the vacuum energy formula in terms of
asymptotic quantities that might be useful to extend the computations to
higher odd dimensions.

\begin{description}
\item[PACS numbers] 11.25.Tq, 04.05.-h, 04.50.Gh, 04.60.Cf.

\end{description}
\end{abstract}

\maketitle





\section{\label{sec:Intro}Introduction.}

In general relativity, black hole solutions play a fundamental role in the
description and the understanding of the gravitational interaction at both
macroscopic and microscopic scales. These objects can be described by a
given set of parameters, namely mass, angular momentum and electric charge.
Our interest lies in Kerr solutions, which have mass and angular momentum,
i.e., are free to rotate.

From the astrophysics point of view, the observable Universe is made of
countless rotating objects, whose geometry is given  in a good approximation
by a Kerr spacetime. Even though the first black hole solution was proposed
in 1916 by Schwarzschild, it took more than 40 years to generalize it to a
rotating case. The main difficulty  for that was the lack of
spherical symmetry and the nondiagonal elements in the metric tensor.

In higher dimensions, the situation is far more complicated. In fact, the
number of rotation parameters increases in one every two additional
dimensions. For this reason, any type of calculation in this geometry quickly turns
very involved as we go up in the spacetime dimension.

Rotating solutions with anti-de Sitter (AdS) asymptotics have become relevant in the context
of anti-de Sitter/conformal field theory (AdS/CFT) correspondence \cite%
{Maldacena,Gubser,Witten}. Examples of holographic studies which consider this type of
solution can be found \cite{Hawkinghunter,Awad1,Awad2,Das}. In all the
cases, the properties of the holographic stress tensor in the gravity side
(e.g., Weyl anomalies) are matched with the ones defined in a boundary CFT
which lives on a rotating Einstein Universe.
In the same vein, further insight on this problem was provided in Refs.\cite{Cai, Papadimitriou, Gibbons2, Awad3}.

Quasinormal modes for rotating solutions were studied in Refs.\cite%
{Moss,Dias}. Perturbations of the black hole metric are equivalent to
the perturbation of a thermal state on the boundary CFT. The time evolution of
the perturbed solution is dual to the time evolution of the thermal state
state fluctuations.

Other theoretical developments that involve Kerr solutions identify
Hawking-Page transitions \cite{Hawkingpage} with confinement/deconfinement
transitions in the strongly coupled regime of the boundary gauge theory \cite%
{Murata:2008bg}. The system also exhibits a gap between the transition lines in
both sides of the duality. In Ref.\cite{Tsai:2011gv} a phase diagram structure
similar to a Weiss ferromagnetic system and a van der Waals liquid/gas
system for certain critical temperature was found. Sonner studied in Ref.%
\cite{Sonner:2009fk} the rotating extension of the idea of holographic
superconductor, in which the model features a superconducting phase
transition on the boundary of a Kerr-Newman black hole.

Nonlinear spinning solutions to fluid mechanics were constructed in Ref.\cite{Bhattacharyya:2007vs, Caldarelli}.
The duality is realized through the identification of the stress
tensor and the thermodynamic quantities in the fluid side with the boundary
stress tensor and thermodynamics of large rotating black holes in the
gravity sector.

Thermodynamic instabilities in Kerr solutions from quantum corrections in
the partition function were found in Refs.\cite%
{Monteiro:2009ke,Monteiro:2009tc,Dias:2009iu}, as the action and the heat
capacity can turn negative.

Extracting holographic quantities from the conformal boundary in AdS
gravity requires one to write down the Einstein equations for an asymptotic
form of the metric \cite{fg}. Solving order by order in powers of the radial
coordinate, the divergences in the variation of the action are canceled by
the addition of local (intrinsic) counterterms . The procedure proves to be
satisfactory in many cases but, however, the method in higher dimensions
becomes far more involved. As a consequence, there is no general expression
for the counterterm series in an arbitrary dimension.
The background independence of the counterterm method is reflected in a nonzero
value for the energy of the AdS vacuum $E_{vac}$ in odd spacetime dimensions \cite{Balasubramanian:1999re}.
The total energy $E=M+E_{vac}$, where $M$ is the Hamiltonian mass, appears
also in thermodynamic relations that involve the evaluation of the Euclidean
action, in whichever method that does not use background subtraction \cite{Emparan:1999pm}. The
zero-point energy for asymptotically AdS (AAdS) black holes can be identified, within the
framework of AdS/CFT, with the Casimir energy of the conformal theory. In
particular, in five dimensions, matching the vacuum energy $%
E_{vac}=3\ell ^{2}/32\pi G$ of Schwarzschild-AdS black holes with the
Casimir energy of $\mathcal{N}$ $=4$ super Yang-Mills theory on the boundary
provided one of the first realizations of the gauge/gravity correspondence.

Similar computations have been performed for rotating AAdS solutions \cite%
{Hawkinghunter,Awad1,Awad2,Das}.

The value of the vacuum energy also plays a role in the proof of positivity
of energy for asymptotically AdS spacetimes in odd dimensions \cite{Horowitz:1998ha,Cheng:2005wk}.

In this article, we extend the results for the vacuum energy in Refs. \cite%
{Hawkinghunter,Awad1,Awad2,Das} to Kerr-AdS black holes with a maximal
number of rotation parameters up to nine spacetime dimensions, using
Kounterterm regularization for AdS gravity \cite{Olea:2006vd} and exploiting
the Kerr-Schild form of metric for rotating solutions.

This paper is organized as follows: in Sec.\ref{sec:KTAdS} we review
Kounterterm regularization for AdS gravity. Once the \emph{extrinsic}
counterterms have been introduced, conserved quantities derived within this
regularization scheme  are revised in Sec.\ref{sec:KTodd}. In particular, we made use of the separability
of the Kounterterm charges in a part that gives the black hole mass and angular
momentum, and another one that produces the vacuum energy. In Sec.\ref{sec:Kerrmetric}, the
metric of Kerr-AdS black holes is cast in Kerr-Schild form, in order to
isolate the part that contributes to the vacuum energy of rotating
solutions. In Sec.\ref{sec:VacKerr}, explicit results up to nine
dimensions are shown. Some properties of the zero-point energy for Kerr-AdS
are discussed. Finally, the last section is devoted to conclusions and
prospects.

\bigskip

\section{\label{sec:KTAdS}Kounterterm regularization in AdS gravity}

For more than a decade, the AdS/CFT correspondence \cite%
{Maldacena,Gubser,Witten} has provided a concrete realization of the
long-standing idea of holographic principle. This form of gauge/gravity
duality has triggered a growing interest in the community, as a useful tool to
describe strongly coupled systems. Maldacena's conjecture has gone beyond
string theory to be applied in areas as diverse as relativistic hydrodynamics, condensed matter, and quantum chromodynamics.

This duality postulates the equivalence between the partition function of
AdS gravity and the one of a boundary CFT, i.e., $Z_{AdS}(\phi)=Z_{CFT}(%
\phi_{0})$. Here,  we understand that the field $\phi$, which lives in
the bulk of the spacetime, takes  the value $\phi_{0}$ as one approaches the
boundary. In the boundary theory, $\phi_{0}$  is interpreted as a source for
a pointlike operator $\mathcal{O}$. In the low-energy limit, the classical
gravitational action can be used to compute the partition function of the
CFT. Also, physical quantities defined in a finite-temperature field theory
can be understood in terms of thermodynamic properties of black holes
in the bulk.

For a suitable realization of the gauge/gravity duality, it is necessary to
render the gravitational action finite. Within this framework, the
convergence of the gravity action for asymptotically AdS spacetimes is
achieved carrying out the holographic renormalization program
\cite{Henningson:1998gx,de Haro:2000xn,Skenderis:2002wp}, which results
in the addition to the original action of local (intrinsic) counterterms \cite{Balasubramanian:1999re,Emparan:1999pm} on
top of a Gibbons-Hawking term \cite{Gibbons4,York}.

Because of the fact that there is no closed formula for the counterterms in an
arbitrary dimension, an alternative series was given in Refs.\cite%
{Olea:2005gb,Olea:2006vd}. This proposal, valid for all dimensions,
considers the addition of boundary terms which are a polynomial in the
extrinsic curvature $K_{i j}$ and the boundary Riemman tensor. By adding
this structure at the boundary, one gets a regularized Euclidean action and
is able to reproduce the correct black hole thermodynamics in AAdS gravity.

Conserved quantities derived within this framework are particularly useful
to deal with solutions with a more complicated structure in the metric, which is the case under
investigation in this paper. This is especially relevant in high enough dimensions,
as we shall discuss below.

\bigskip

Let us take the action for Einstein gravity with negative cosmological
constant in $D=d+1$\ dimensions

\begin{equation}
I=-\frac{1}{16\pi G}\int_{M}d^{d+1}x\sqrt{-g}(R-2\Lambda
)+c_{d}\int_{\partial M}d^{d}xB_{d} ,
\end{equation} where $\Lambda =-\frac{d\left( d-1\right) }{2\ell ^{2}}$ and $R$ is the
spacetime Ricci scalar. The boundary term $B_{d}$ is added for the purpose of
finiteness of the conserved quantities and, at the same time, it produces a
well-posed action principle for AAdS spacetimes.

The spacetime geometry can be described in terms of Gaussian coordinates
\begin{equation}
ds^{2}=g_{\mu \nu }dx^{\mu }dx^{\nu }=N^{2}(r)dr^{2}+h_{ij}(r,x)dx^{i}dx^{j},
\label{Gauss radial}
\end{equation}%
where $r$ is the radial coordinate. The manifold $M$ possesses a single
boundary $\partial M$, which is located at radial infinity. Indeed, for $%
r=\infty $, the metric $h_{ij}$ accounts for the intrinsic properties of the
boundary, which is parametrized by the coordinates $\{x^{i}\}$. In turn, the
extrinsic properties are given in terms of an outward-pointing spacelike unit
normal $n_{\mu }=(n_{r},n_{i})=(N,\vec{0})$, as the extrinsic curvature is
defined as the Lie derivative of the boundary metric along $n$, that is,
\begin{equation}
K_{ij}=\mathcal{L}_{n}h_{ij}\,.
\end{equation}%
In a more explicit form, the extrinsic curvature is given by
\begin{equation}
K_{ij}:=-{\frac{1}{{2N}}}h_{ij}^{\prime }\,,
\end{equation}%
where prime denotes radial derivative.

For the case of the odd dimensions ($D=2n+1$) we are interested in, the boundary
term adopts a compact form when expressed with the help of two parametric
integrations
\begin{eqnarray}
B_{2n}=2n\sqrt{-h}\int_{0}^{1}du\int_{0}^{u}ds\ \delta _{\lbrack
i_{1}...i_{2n}]}^{[j_{1}...j_{2n}]}K_{j_{1}}^{i_{1}}&& \delta
_{j_{2}}^{i_{2}} \left( \frac{1}{2}\mathcal{R}%
_{j_{3}j_{4}}^{i_{3}i_{4}}-u^{2}K_{j_{3}}^{i_{3}}K_{j_{4}}^{i_{4}}+\frac{%
s^{2}}{\ell ^{2}}\delta _{j_{3}}^{i_{3}}\delta _{j_{4}}^{i_{4}}\right)
\times \cdots  \notag \\
 \cdots \times &&\left( \frac{1}{2}\mathcal{R}%
_{j_{2n-1}j_{2n}}^{i_{2n-1}i_{2n}}-u^{2}K_{j_{2n-1}}^{i_{2n-1}}K_{j_{2n}}^{i_{2n}}+%
\frac{s^{2}}{\ell ^{2}}\delta _{j_{2n-1}}^{i_{2n-1}}\delta
_{j_{2n}}^{i_{2n}}\right),
\end{eqnarray}%
where $h$ is the determinant of the boundary metric , $\mathcal{R}_{kl}^{ij}$
is the boundary Riemman tensor and $\delta _{\lbrack
i_{1}...i_{2n}]}^{[j_{1}...j_{2n}]}$ is a totally antisymmetric product of $%
2n$ Kronecker deltas (for conventions, see Appendix \ref{Delta}). When
expanded, the above expression can be seen as a polynomial in the extrinsic
and intrinsic curvatures, where the coefficients are obtained once the
integrations in $s$ and $u$ are performed. In that respect, the use of
parametric integrations is not a mere formality, but provides an operational
tool to derive general expressions for the conserved quantities, regardless of
the spacetime dimension. In particular, Kounterterm regularization leads to
the only formula for the vacuum energy for AAdS spaces in all odd
dimensions, which is covariant with respect to the boundary metric $h_{ij}$.

The finiteness of the conserved quantities is achieved once the coupling
constant of $B_{2n}$ is chosen as
\begin{equation}
c_{2n}:=\frac{1}{16\pi G}\frac{(-1)^{n}\ell ^{2n-2}}{2^{2n-2}n(n-1)!^{2}}.
\label{c2n}
\end{equation}%
The above value is singled out by the cancelation of leading-order terms in
the asymptotic expansion of the surface term produced by the variation of
the action. It is a remarkable fact that this choice of $c_{2n}$ also
eliminates the rest of divergences that appear in the surface term and,
subsequently, in the conserved charges. There is no other explanation for this
property, other than saying that in Kounterterm regularization the boundary
terms are related to well-known mathematical structures as topological
invariants and Chern-Simons densities. Therefore, it is hard to think of a
more $\emph{geometric}$ object that can be added to the gravity action for
the purpose of regularization.

In addition, there is a partial proof (in even spacetime dimensions) that
Kounterterms are able to generate the standard counterterm series upon a
suitable expansion of the extrinsic curvature \cite{Miskovic:2009bm}. This
conclusion is quite remarkable: holographic renormalization is equivalent to
the addition of topological invariants in even dimensions. This also means
that the \emph{extrinsic} regularization scheme can be converted into an
\emph{intrinsic} one, which is necessary to recover standard holographic
quantities in AAdS gravity.

\section{\label{sec:KTodd}Kounterterm charges in odd dimensions}

Provided the boundary coincides with the asymptotic region, the Noether
procedure leads to conserved quantities associated to a set of asymptotic
Killing vectors $\{\xi ^{i}\}$. The conservation of the Noether current $%
\partial _{\mu }J^{\mu }=0$ implies that $J$ can be written locally as a
total derivative which, in turn, implies the existence of a conserved quantity%
\begin{equation}
Q\left[ \xi \right] =\int\limits_{\partial M}d^{d}x\,\frac{1}{N}\,n_{\mu
}J^{\mu }(\xi )\,.  \label{Qxi1}
\end{equation}%
Altogether, it was shown in Ref.\cite{Olea:2006vd} that the radial component of
the Noether current $J^{r}=\frac{1}{N}\,n_{\mu }J^{\mu }$ is globally a
total derivative on $\partial M$, that is,
\begin{equation}
Q\left[ \xi \right] =\int\limits_{\partial M}d^{d}x\,\partial _{j}\left(
\sqrt{-h}\,\xi ^{i}\,\left( q_{i}^{j}+q_{(0)i}^{j}\right) \right) \,.
\end{equation}%
Using the Stokes theorem, the above quantity can be written as a surface
integral in $(d-1)$ dimensions. In order to do so, we foliate the boundary $%
\partial M$ in Arnowitt-Deser-Misner form with the coordinates $x^{i}=(t,y^{m})$%
\begin{eqnarray}
h_{ij}dx^{i}dx^{j} &=&-{\tilde{N}}^{2}(t)dt^{2}+\sigma _{mn}\left( dy^{m}+{%
\tilde{N}}^{m}dt\right) \left( dy^{n}+{\tilde{N}}^{n}dt\right) ,   \label{ADM boundary} \\
&&\sqrt{-h}=\tilde{N}\sqrt{\sigma }.  \label{rooth}
\end{eqnarray}%
The lapse function ${\tilde{N}}$ appears in the timelike normal $u_{i}$ (that
generates the foliation) as $u_{i}=(u_{t},u_{m})=(-\tilde{N},\vec{0})$. The
tensor $\sigma _{mn}$ represents the metric of the spatial section at
constant time. We denote this surface by the symbol $\Sigma _{\infty }$.

The Noether charge in odd spacetime dimensions is expressed as the sum of two parts
\begin{equation}
Q[\xi ]=q[\xi ]+q_{(0)}[\xi ]  \label{Qxi total} ,
\end{equation}%
where the first integral
\begin{equation}
q[\xi ]=\int\limits_{\Sigma _{\infty }}d^{2n-1}y\sqrt{\sigma }\,u_{j}\,q_{i}^{j}\xi
^{i} , \label{qxi}
\end{equation}%
produces the mass and angular momentum for AAdS spacetimes, with an
integrand given by%
\begin{eqnarray}
q_{i}^{j} &=&-{\frac{1}{2^{n-2}}}\delta _{\lbrack k\;i_{1}\cdots
i_{2n-1}]}^{[j\;j_{1}\cdots j_{2n-1}]}K_{i}^{k}\delta _{j_{1}}^{i_{1}}\left[
{\frac{1}{{16\pi G(2n-1)!}}}\delta _{\lbrack
j_{2}\,j_{3}]}^{[i_{2}\,i_{3}]}\times \cdots \times \delta _{\lbrack
j_{2n-2}\,j_{2n-1}]}^{[i_{2n-2}\,i_{2n-1}]}\right.  \notag \\
&&\left. +n\,c_{2n}\int\limits_{0}^{1}du\left( R_{j_{2}j_{3}}^{i_{2}\,i_{3}}+%
{\frac{u^{2}}{\ell ^{2}}}\delta _{\lbrack
j_{2}\,j_{3}]}^{[i_{2}i_{3}]}\right) \times \cdots \times \left(
R_{j_{2n-2}j_{2n-1}}^{i_{2n-2}i_{2n-1}}+{\frac{u^{2}}{\ell ^{2}}}\delta
_{\lbrack j_{2n-2}j_{2n-1}]}^{[i_{2n-2}i_{2n-1}]}\right) \right].  \label{qij}
\end{eqnarray}

On the other hand, the second part

\begin{equation}
q_{(0)}[\xi ]=\int\limits_{\Sigma _{\infty }}d^{2n-1}y\sqrt{\sigma }\,%
u_{j}\,q_{(0)i}^{j}\xi ^{i} ,  \label{q0xi}
\end{equation}%
is given in terms of the tensor%
\begin{eqnarray}
q_{(0)i}^{j}=2n\,c_{2n}\delta _{\lbrack k\,i_{1}\cdots
i_{2n-1}]}^{[j\,j_{1}\cdots j_{2n-1}]}\int\limits_{0}^{1}&& du\,u\, \left(
K_{i}^{k}\delta _{j_{1}}^{i_{1}}+K_{j_{1}}^{k}\delta _{i}^{i_{1}}\right)
\left( \frac{1}{2}\mathcal{R}%
_{j_{2}j_{3}}^{i_{2}i_{3}}-u^{2}K_{j_{2}}^{i_{2}}K_{j_{3}}^{i_{3}}+{\frac{%
u^{2}}{\ell ^{2}}}\delta _{j_{2}}^{i_{2}}\delta _{j_{3}}^{i_{3}}\right)
\times \cdots  \notag  \label{cv} \\
&&\cdots \times \left( \frac{1}{2}\mathcal{R}%
_{j_{2n-2}j_{2n-1}}^{i_{2n-2}i_{2n-1}}-u^{2}K_{j_{2n-2}}^{i_{2n-2}}K_{j_{2n-1}}^{i_{2n-1}}+%
{\frac{u^{2}}{\ell ^{2}}}\delta _{j_{2n-2}}^{i_{2n-2}}\delta
_{j_{2n-1}}^{i_{2n-1}}\right) .  \label{q0ij}
\end{eqnarray}

Properties of the above formulas are extensively employed in Sec.\ref{sec:VacKerr}, where we compute
the vacuum energy for Kerr-AdS metric in odd spacetime dimensions.

In the next section, we review the construction of the Kerr-AdS metric in
$D=2n+1$ dimensions.

\section{\label{sec:Kerrmetric}Kerr black hole metric in odd spacetime dimensions.}

The asymptotically flat Kerr spacetime in four dimensions \cite{Kerr} can be
obtained as a perturbation to the Minkowski metric that is linear in the
parameter $M$ (related to the black hole mass) \cite{KerrSchild}. The
resulting line element adopts the form

\begin{equation}
ds^{2}=\eta _{\mu \nu }dx^{\mu }dx^{\nu }+{\frac{2M}{{U}}}(k_{\mu }dx^{\mu
})^{2}\,,
\end{equation}%
where $k_{\mu }$ is a null geodesic vector for the seed metric $\eta _{\mu
\nu }$ as for the full metric $g_{\mu \nu }$, and the function $U$ is given
by%
\begin{equation}
U=r+{\frac{a^{2}z^{2}}{{r^{3}}}\,,}
\end{equation}%
where $r=(0,\infty )$, $z$ is the axis around which the rotation will be
defined and $a$ is a parameter that represents the squashing of the sphere.

In Cartesian coordinates, the explicit form of the deformation $k=k_{\mu
}dx^{\mu }$ is
\begin{equation}
k=dt+{\frac{r(x\,dx+y\,dy)+a(x\,dy-y\,dx)}{{r^{2}+a^{2}}}}+{\frac{z\,dz}{{r}}%
} ,
\end{equation}%
and $r$ is defined by the ellipsoidal hypersurface%
\begin{equation}
{\frac{x^{2}+y^{2}}{{r^{2}+a^{2}}}}+{\frac{z^{2}}{{r^{2}}}}=1\,.
\end{equation}

In a similar way, if one includes a cosmological constant $\Lambda $ into
the gravitational action, the metric takes a linearized form around de
Sitter or anti-de Sitter background $\bar{g}_{\mu \nu }$ (see, e.g., Ref.\cite{Carter}),
\begin{equation}
g_{\mu \nu }=\bar{g}_{\mu \nu }+{\frac{2M}{U}}k_{\mu }k_{\nu }\,.
\label{Kerr-Schild AdS}
\end{equation}%
Once again, $k_{\mu }$ is a null vector for both metrics.

The generalization of the Kerr metric to higher dimensions makes use of the same
properties seen above \cite{MyersPerry,Gibbons:2004uw,Gibbons:2004ai}.

In particular, in odd spacetime dimensions ($D=2n+1$) we consider the parametrization
for the unit sphere $S^{2n-1}$, which considers $n$ two-planes, with $2n$ coordinates subjected to a
constraint. The coordinates of the $i$th plane are $(u_{i},v_{i})$, whose
polar form is given by
\begin{equation}
u_{i}+iv_{i}=\hat{\mu}_{i}e^{i\phi _{i}} ,  \label{u+iv}
\end{equation}%
where we have introduced $n$ azimuthal angles $\phi _{i}$ and $n$ direction
cosines $\hat{\mu}_{i}$, which satisfy the constraint
\begin{equation}
\sum_{i=1}^{n}{\hat{\mu}_{i}^{2}}=1.
\end{equation}

Adding up the time and radial directions in the line element, the global AdS
metric in this coordinate set adopts the form
\begin{equation}
d\bar{s}^{2}=-\left( 1+{\frac{y^{2}}{{l^{2}}}}\right) dt^{2}+{\frac{dy^{2}}{{%
1+{\frac{y^{2}}{{l^{2}}}}}}} +y^{2}\sum_{i=1}^{n}\left( d\hat{\mu}_{i}^{2}+%
\hat{\mu}_{i}^{2}d\phi _{i}^{2}\right) .  \label{AdS global mu}
\end{equation}%
If the dimension of the spacetime is $d+1$, the number $N$ of independent rotation parameters
$\{a_{i}\}$ corresponds to the number of Casimir invariants of $SO(d)$,
that is, $N=\left[ \frac{d}{2}\right] $.

Then, we pass from the sphere parametrization in Eq.(\ref{u+iv}) to a new
set of spheroidal coordinates defined by the transformation
\begin{equation}
\left( 1-{\frac{a_{i}^{2}}{\ell {^{2}}}}\right) y^{2}\hat{\mu}%
_{i}^{2}=\left( r^{2}+a_{i}^{2}\right) \mu _{i}^{2} ,
\end{equation}%
where, once again, the variable $\mu _{i}$ is constrained by the equation
\begin{equation}
\sum_{i}\mu _{i}^{2}=1.
\end{equation}

With this transformation replaced in the metric of global AdS, we can
express the line element in terms of the new variables $(r,\mu _{i})$, such
that the vacuum spacetime (\ref{AdS global mu}) is written as%
\begin{eqnarray}
d\bar{s}^{2}=- &&W\left( 1+{\frac{r^{2}}{{\ \ell ^{2}}}}\right)
dt^{2}+Fdr^{2}+\sum_{i=1}^{n}{\frac{r^{2}+a_{i}^{2}}{\Xi _{i}}}\left( d\mu
_{i}^{2}+\mu _{i}^{2}d\phi _{i}^{2}\right) +  \notag \\
&&-{\frac{1}{{W}\left( {1+{\frac{r^{2}}{\ell {^{2}}}}}\right) \ell {^{2}}}}%
\left( \sum_{i=1}^{n}{\frac{\left( r^{2}+a_{i}^{2}\right) \mu _{i}d\mu _{i}}{%
\Xi _{i}}}\right) ^{2} ,  \label{dsbar22}
\end{eqnarray}%
where
\begin{equation}
\Xi _{i}={1-{\frac{a_{i}^{2}}{\ell {^{2}}}\,.}}  \label{Xi}
\end{equation}%
The functions $W$ and $F$ that appear in the metric are given by
\begin{equation}
W\equiv \sum_{i=1}^{n}{\frac{\mu _{i}^{2}}{\Xi _{i}}},\qquad F\equiv {\frac{%
r^{2}}{{1+{\frac{r^{2}}{\ell {^{2}}}}}}}\sum_{i=1}^{n}{\frac{\mu _{i}^{2}}{{%
r^{2}+a_{i}^{2}}}}.  \label{WandF}
\end{equation}%
This metric obtained as the deformation of global AdS geometry will be used
to find the vacuum energy of Kerr-AdS in the next section. This is justified
by the fact that the black hole mass $M$ does not appear in $d\bar{s}^{2}$
but only in the full metric as $ds^{2}=d\bar{s}^{2}+{\frac{2M}{{U}}}(k_{\mu
}dx^{\mu })^{2}$.

Finally, the explicit form of the perturbation to the deformed vacuum metric
$d\bar{s}^{2}$ is
\begin{equation}
k_{\mu }dx^{\mu }=Wdt+Fdr-\sum_{i=1}^{n}{\frac{a_{i}\mu _{i}^{2}}{\Xi _{i}}}%
d\phi _{i} ,
\end{equation}%
and where
\begin{equation}
U=\sum_{i=1}^{n}{\frac{\mu _{i}^{2}}{{r^{2}+a_{i}^{2}}}}%
\prod_{j=1}^{n}(r^{2}+a_{i}^{2}).
\end{equation}%
The full Kerr-AdS metric is usually expressed in terms of Boyer-Lindquist
coordinates which eliminate the components $g_{\mu r}$ with $\mu \neq r$,
i.e., no cross terms between $dr$ and the other coordinate differentials.
Indeed, it would be convenient, putting the metric in the Gaussian form (\ref%
{Gauss radial}), e.g., to evaluate the mass and angular momenta \cite{Deser:2005jf} for rotating
black holes from Eqs.(\ref{qxi}),(\ref{qij}). However, in the next section it
is argued that, for the purpose of vacuum energy computation it is enough to
consider just the deformation induced by the rotation parameters on the
global AdS spacetime, i.e., the line element (\ref{dsbar22}).

In addition, Kerr-AdS metric can be expressed in Kerr-Schild form,  which
splits it in two sectors. The first one is the rotating version of global
AdS space and the second, a part proportional to the mass parameter. In this
way, we can be sure that there will not be missing contributions to the
vacuum  energy once we switch off the mass. This justifies the fact that, in
order to perform the calculations relevant for this paper, we can restrict
ourselves to the rotating  global AdS metric.

\section{\label{sec:VacKerr}Vacuum energy in Kerr-AdS.}

The deformation of the AdS vacuum defined by Eqs. (\ref{dsbar22}) and (\ref%
{WandF}) preserves the constant-curvature property of global AdS spacetime.
This means that the global transformations performed in order to obtain the
metric (\ref{dsbar22}) do not modify the local condition%
\begin{equation}
R_{\mu \nu }^{\alpha \beta }+\frac{1}{\ell ^{2}}\delta _{\lbrack \mu \nu
]}^{[\alpha \beta ]}=0.  \label{LAdS bar}
\end{equation}%
From the argument that follows it is evident that the conserved quantities
associated to this part of the metric are identically zero. Indeed, it can
be shown that the part $q(\xi )$ of the total charge (\ref{Qxi total}) that
produces the mass and angular momentum for AAdS black holes can have its
integrand factorized as
\begin{equation}
q_{i}^{j}=\frac{nc_{2n}}{2^{n-2}}\delta _{\lbrack
i_{1}i_{2}...i_{2n}]}^{[jj_{2}...j_{2n}]}K_{i}^{i_{1}}\delta
_{j_{2}}^{i_{2}}\left( R_{j_{3}j_{4}}^{i_{3}i_{4}}+\frac{1}{\ell ^{2}}\delta
_{\lbrack j_{3}j_{4}]}^{[i_{3}i_{4}]}\right) \mathcal{P}%
_{j_{5}...j_{2n}}^{i_{5}...i_{2n}}\left( R,\delta \right) .
\label{qijfactor}
\end{equation}%
Here, $\mathcal{P}\left( R,\delta \right) $ is a polynomial of $(n-2)$
degree in the spacetime Riemann tensor $R_{kl}^{ij}$ (its projection at the
boundary) and the antisymmetrized Kronecker delta $\delta _{\lbrack
kl]}^{[ij]}$

\begin{eqnarray}
\mathcal{P}_{j_{5}...j_{2n}}^{i_{5}...i_{2n}} \left( R,\delta \right)
=\sum_{p=0}^{n-2}\frac{D_{p}}{\ell ^{2p}}%
R_{j_{5}j_{6}}^{i_{5}i_{6}}...R_{j_{2(n-p)-1}j_{2(n-p)}}^{i_{2(n-p)-1}i_{2(n-p)}} \delta _{\lbrack j_{2(n-p)+1}j_{2(n-p+1)}]}^{[i_{2(n-p)+1}i_{2(n-p+1)}]}...\delta _{\lbrack j_{2n-1}j_{2n}]}^{[i_{2n-1}i_{2n}]},
\end{eqnarray}%
with the coefficients of the expansion given by

\begin{equation}
D_{p}=\sum_{q=0}^{p}\frac{(-1)^{p-q}}{2q+1}{\binom{{n-1} }{{q} }}.
\end{equation}%
Therefore, any space satisfying the condition (\ref{LAdS bar}) globally will
posses vanishing charges.

From the explicit form of the full metric, we can notice that $M$ does not
appear in $d\bar{s}^{2}$ in Eq.(\ref{Kerr-Schild AdS}). That means that the
parameter $M$ cannot affect the value of the vacuum energy (which is obvious
when we think that the vacuum state corresponds to a vanishing mass).

On the other hand, it can be seen that the electric part of the Weyl tensor $%
\mathcal{E}_{j}^{i}\sim n_{\mu }n^{\nu }W_{\nu j}^{\mu i}$ of the full
Kerr-AdS metric is always proportional to $M$. In this way, it correctly
reproduces the mass and angular momentum from the Ashtekar-Magnon-Das charge
definition for AAdS \cite{AM,AD,Das}. Furthermore, the Weyl tensor is
--on-shell-- proportional to the right-hand side of Eq.(\ref{LAdS bar}) and,
therefore, $M$ should not enter into the expression of $q_{(0)}$.

In summary, we only need the sector in the metric that corresponds to the
deformation of global AdS spacetime in order to compute the zero-point
energy (\ref{q0xi}). As the above integral is defined in the limit for $%
r\rightarrow \infty $, we shall consider the asymptotic expansion of the
intrinsic and extrinsic curvatures.

Taking the metric of global AdS in oblate coordinates in Eqs. (\ref{dsbar22}%
) and (\ref{WandF}) and writing down the direction cosines in terms of polar
angles (see the Appendix \ref{apC}), we see that the squared root of the
determinant of the boundary metric behaves as
\begin{equation}
\sqrt{-h}=\tilde{N}\sqrt{\sigma }\sim r^{2n}+\mathcal{O}(r^{2n-2}),
\end{equation}%
where the function $\tilde{N}$ appears in the Arnowitt-Deses-Misner foliation (\ref{ADM
boundary}).

From explicit computations in the oblate-AdS sector of the Kerr-AdS metric in an
arbitrary dimension, one can see that the asymptotic expansion of the
extrinsic curvature is
\begin{equation}
K_{j}^{i}=-\frac{\delta _{j}^{i}}{\ell }+\frac{\ell A_{j}^{i}(\theta ,\phi )%
}{r^{2}}+\mathcal{O}\left( \frac{1}{r^{4}}\right) ,
\end{equation}%
whereas the intrinsic curvature behaves as
\begin{equation}
\mathcal{R}_{kl}^{ij}\sim \frac{\mathcal{B}_{kl}^{ij}(\theta ,\phi )}{r^{2}}+%
\mathcal{O}\left( \frac{1}{r^{4}}\right) , \label{expansionRkerr}
\end{equation}%
where $A_{j}^{i}$ and $\mathcal{B}_{kl}^{ij}$ are  tensor coefficients
which do not have radial dependence.

In the expression (\ref{q0ij}), we have $(n-1)$ terms of the form
\begin{equation}
\mathcal{R}_{kl}^{ij}-u^{2}(K_{k}^{i}k_{l}^{j}-K_{l}^{i}K_{k}^{j})+\frac{%
u^{2}}{\ell ^{2}}\delta _{\lbrack kl]}^{[ij]}\sim \frac{1}{r^{2}}[\mathcal{B}%
_{kl}^{ij}+u^{2}(\delta _{k}^{i}A_{l}^{j}+\delta _{l}^{j}A_{k}^{i}-\delta
_{l}^{i}A_{k}^{j}-\delta _{k}^{j}A_{l}^{i})]+\mathcal{O}\left( \frac{1}{r^{4}%
}\right). \label{factorKerrAdS}
\end{equation}%
In general, the integration in the continuous parameter $u$ present in the
formula for vacuum energy is quite complicated to solve.

However, in any dimension, from explicit computations in the Kerr-AdS metric,
one can notice that the leading-order term in the expansion of the intrinsic
curvature \ is the skew-symmetric product of $A_{j}^{i}$ with a Kronecker
delta, that is,
\begin{equation}
\mathcal{B}_{kl}^{ij}=-(\delta _{k}^{i}A_{l}^{j}+\delta
_{l}^{j}A_{k}^{i}-\delta _{l}^{i}A_{k}^{j}-\delta _{k}^{j}A_{l}^{i}).
\end{equation}%
The above reasoning allows us to factorize the expression (\ref{factorKerrAdS}) in terms of the next-to-leading order in the expansion of the
extrinsic curvature as
\begin{equation}
\mathcal{R}_{kl}^{ij}-u^{2}(K_{k}^{i}k_{l}^{j}-K_{l}^{i}K_{k}^{j})+\frac{%
u^{2}}{\ell ^{2}}\delta _{\lbrack kl]}^{[ij]}\sim \frac{(u^{2}-1)}{r^{2}}%
(\delta _{k}^{i}A_{l}^{j}+\delta _{l}^{j}A_{k}^{i}-\delta
_{l}^{i}A_{k}^{j}-\delta _{k}^{j}A_{l}^{i})+\mathcal{O}\left( \frac{1}{r^{4}}%
\right) .
\end{equation}%
The formula for the zero-point energy (\ref{q0ij}) also involves the
combination
\begin{equation}
K_{i}^{k}\delta _{j_{1}}^{i_{1}}+K_{j_{1}}^{k}\delta _{i}^{i_{1}}=-\frac{1}{%
\ell }\left( \delta _{i}^{k}\delta _{j_{1}}^{i_{1}}+\delta
_{j_{1}}^{k}\delta _{i}^{i_{1}}\right) +\frac{\ell }{r^{2}}\left(
A_{i}^{k}\delta _{j_{1}}^{i_{1}}+A_{j_{1}}^{k}\delta _{i}^{i_{1}}\right) +%
\mathcal{O}\left( \frac{1}{r^{4}}\right),
\end{equation}%
which, when multiplied by the totally antisymmetric Kronecker delta,
produces the identical cancelation of its first term. Just by a simple
power-counting argument in the radial coordinate, the formula of the vacuum
energy for Kerr-AdS reduces to

\begin{equation}
q_{(0)i}^{j}=(-2)^{n-1}c_{2n}\,\delta _{\lbrack k\,i_{1}\cdots
i_{2n-1}]}^{[j\,j_{1}\cdots j_{2n-1}]}\left( A_{i}^{k}\delta
_{j_{1}}^{i_{1}}+A_{j_{1}}^{k}\delta _{i}^{i_{1}}\right)
A_{j_{2}}^{i_{2}}\delta _{j_{3}}^{i_{3}}\times \cdots \times
A_{j_{2n-2}}^{i_{2n-2}}\delta _{j_{2n-1}}^{i_{2n-1}},
\end{equation}%
after performing a trivial integration in the parameter $u$.

Then, the vacuum energy formula (\ref{q0xi}) in the limit $r\rightarrow
\infty $ is written as
\begin{eqnarray}
E_{vac} &=&-\int\limits_{\Sigma _{\infty }}d^{2n-1}y\,\sqrt{-h}\,q_{(0)t}^{t} ,
\notag \\
&=&-{\frac{\ell ^{2n-1}}{2^{n+3}\pi Gn!}}\delta _{\lbrack p_{1}p_{2}\cdots
p_{n}]}^{[n_{1}n_{2}\cdots n_{n}]}\int\limits_{\Sigma _{\infty }}d^{2n-1}y%
\sqrt{-h}\frac{1}{r^{2n}}\left( A_{n_{1}}^{p_{1}}-A_{t}^{t}\,\delta
_{n_{1}}^{p_{1}}\right) \,A_{n_{2}}^{p_{2}}\times \cdots \times
A_{n_{n}}^{p_{n}} ,  \label{Evacang}
\end{eqnarray}%
in terms of the of the next-to-leading order quantities in the expansion of
both the extrinsic and intrinsic curvatures. \ Here, the indices $%
\{n_{i},p_{i}\}$ are restricted to the angular part of the boundary metric,
that is, the angles of the sphere $S^{2n-1}$ (see Appendix \ref{apC}).
Explicit results up to nine dimensions are given below \cite{GRTensorII}.

\textbf{Five dimensions. }As a warm-up we evaluate the five-dimensional
version of the rotating AdS vacuum spacetime (\ref{dsbar22}-\ref{WandF}),
and we obtain

\begin{equation}
E_{vac}^{(5)}=\frac{3\pi \ell ^{2}}{32G}\left( 1+\frac{\left( \Xi _{a}-\Xi
_{b}\right) ^{2}}{9\Xi _{a}\Xi _{b}}\right) , \label{E05}
\end{equation}%
which is already a standard result in the literature \cite{Awad2,
Papadimitriou,Gibbons2}.

We stress the fact that $E_{vac}$ reduces to the one of a static black hole
with $R\times S^{3}$ topology at the boundary, either when the rotation
parameters vanish or when they equal ($a=b$). As we shall show below, this
feature is also present in higher odd-dimensional Kerr-AdS black holes.

\textbf{Seven dimensions. }The sector with $M=0$ of the Kerr-AdS metric in
seven dimensions considers the deformation of global AdS spacetime by the
action of three rotation parameters. The formula for the vacuum energy, Eqs.(\ref{q0xi}%
) and (\ref{q0ij}), produces

\begin{eqnarray}
E_{vac}^{(7)}&& =-\frac{5\pi ^{2}\ell ^{4}}{128G}  \left(1 +\frac{1}{50\Xi_{a}\Xi_{b}\Xi_{c}}%
\left( (\Xi_{a}-\Xi_{b})(\Xi_{a}-\Xi_{c})(3%
\Xi_{b}+3\Xi_{c}-\Xi_{a}) +\right.\right. \notag \\
&&\Big. +  (\Xi_{b}-\Xi_{c})(\Xi_{b}-\Xi_{a})(3\Xi_{c}+3%
\Xi_{a}-\Xi_{b})+(\Xi_{c}-\Xi_{a})(\Xi_{c}-\Xi_{b})(3\Xi_{a}+3\Xi_{b}-%
\Xi_{c})  \Big).   \label{E07}
\end{eqnarray}

The above expression for the zero-point energy can be rewritten in a more compact way as
\begin{equation}
E_{vac}^{(7)} =-\frac{5\pi ^{2}\ell ^{4}}{128G}  \left(1 +\frac{1}{100\prod\limits_{l}\Xi _{l}}%
\sum_{i}\sum_{j}\sum_{k \neq j} (\Xi_{i}-\Xi_{j})(\Xi_{i}-\Xi_{k})(3\Xi-4\Xi_{i}) \right), \label{E07new}
\end{equation}
where $\Xi=\sum_{l} \Xi_{l}$.

In absence of previous results in the literature to compare with, we take the single-parameter
limit in the above expression (Myers-Perry)

\begin{equation}
E_{vac}^{(7)}=-\frac{\pi ^{2}}{1280\ell ^{2}G\left( 1-\frac{a^{2}}{\ell ^{2}}%
\right) }\left( 50\ell ^{6}-50\ell ^{4}a^{2}+5\ell ^{2}a^{4}+a^{6}\right) ,
\label{E07one}
\end{equation}%
from where we see that it coincides with the value computed using a quasilocal
stress tensor --properly regularized using counterterm method-- by Das and
Mann \cite{Das}, and Awad and Johnson \cite{Awad2}.

It is clear that the vacuum energy for Schwarzschild-AdS \cite{E07}
\begin{equation}
E_{vac}=-\frac{5\pi ^{2}\ell ^{4}}{128G} , \label{E07SchAdS}
\end{equation}%
is degenerated  because it is the same for seven-dimensional Kerr-AdS with
all rotation parameters equal, i.e., $\Xi _{a}=\Xi _{b}=\Xi _{c}$.

\textbf{Nine dimensions. }Evaluating the expression of the vacuum energy for
a nine-dimensional black hole with maximal number of rotation parameters is
quite demanding from a computational point of view. Because the formula
involves a totally anti-symmetric Kronecker delta, the number of calculations
increases drastically with the dimension. However, from the discussion
above, we notice that Eq.(\ref{Evacang}) must be integrated only in angular
variables, reducing the problem in two dimensions with respect to the one of the
spacetime. The result for the zero-point energy in Kerr-AdS in nine
dimensions is then given by%
\begin{eqnarray}
E_{vac}^{(9)} &=&\frac{\pi ^{3}\ell ^{6}}{322560G\Xi _{a}\Xi _{b}\Xi _{c}\Xi
_{d}}\left( 15\Xi _{a}^{4}+15\Xi _{b}^{4}+15\Xi _{c}^{4}+15\Xi
_{d}^{4}-55\Xi _{a}^{3}\Xi _{b}-55\Xi _{a}^{3}\Xi _{c}-55\Xi _{a}^{3}\Xi_{d} \right.   \notag \\
&-&55\Xi _{b}^{3}\Xi _{a}-55\Xi _{b}^{3}\Xi _{c}-55\Xi _{b}^{3}\Xi _{d}-55\Xi _{c}^{3}\Xi
_{a}-55\Xi _{c}^{3}\Xi _{b}-55\Xi _{c}^{3}\Xi _{d}-55\Xi _{d}^{3}\Xi
_{a}-55\Xi _{d}^{3}\Xi _{b} \notag \\
&+&55\Xi _{d}^{3}\Xi _{c}+211\Xi _{a}^{2}\Xi
_{b}\Xi _{c}+211\Xi _{a}^{2}\Xi _{b}\Xi _{d}+211\Xi _{a}^{2}\Xi _{c}\Xi _{d}+211\Xi
_{b}^{2}\Xi _{a}\Xi _{c}+211\Xi _{b}^{2}\Xi _{a}\Xi _{d} \notag \\
&+&211\Xi _{b}^{2}\Xi
_{c}\Xi _{d}+211\Xi _{c}^{2}\Xi _{a}\Xi _{b}+211\Xi _{c}^{2}\Xi _{a}\Xi _{c}+%
211\Xi _{c}^{2}\Xi _{b}\Xi _{d}+211\Xi _{d}^{2}\Xi _{a}\Xi _{b}+211\Xi
_{d}^{2}\Xi _{a}\Xi _{c} \notag \\
&+&211\Xi _{d}^{2}\Xi _{b}\Xi _{c}+29\Xi _{a}^{2}\Xi
_{b}^{2}+29\Xi _{a}^{2}\Xi _{c}^{2}+29\Xi _{a}^{2}\Xi _{d}^{2}+29\Xi
_{b}^{2}\Xi _{c}^{2}+29\Xi _{b}^{2}\Xi _{d}^{2}+29\Xi _{c}^{2}\Xi _{d}^{2}  \notag \\
&+&1569\Xi _{a}\Xi
_{b}\Xi _{c}\Xi _{d}).  \label{Evac9abcd}
\end{eqnarray}%
The reader may check, in a straightforward way, that for the case $a=b=c=d$,
the above expression has the same property as in five and seven dimensions,
as it reduces to the vacuum energy of static spherical black hole%
\begin{equation}
E_{vac}^{(9)}=\frac{35\pi ^{3}\ell ^{6}}{3072G}\,\ .
\end{equation}%
It is evident that there is an equivalent form to Eq.(\ref{Evac9abcd}) that
makes this feature more manifest. Indeed, using all the symmetries under the
exchange of rotation parameters, the vacuum energy can be written as{\
\begin{eqnarray}
E_{vac}^{(9)} &=&\frac{35\pi ^{3}\ell ^{6}}{3072G}\left[ 1+\frac{1}{%
176400\prod\limits_{l}\Xi _{l}}\sum_{i}\sum_{j}\sum_{k \neq j}(\Xi _{i}-\Xi
_{j})(\Xi _{i}-\Xi _{k})\times \right. \notag \\
&&  \times \left( 120\Xi _{i}^{2}-366\left( \Xi _{j}^{2}+\Xi
_{k}^{2}\right) +(\Xi _{j}+\Xi _{k})\left( -2646\Xi _{i}+2106(\Xi -\Xi
_{j}-\Xi _{k})\right) \right) \Big] ,
\end{eqnarray}%

When we take the limit of a single-parameter rotating black hole ($b=c=d=0$%
), $E_{vac}$ adopts the form

\begin{equation}
E_{vac}^{(9)}=\frac{\pi ^{3}}{21504\ell ^{2}G\left( 1-\frac{a^{2}}{\ell ^{2}}%
\right) }\left( 245\ell ^{8}-245\ell ^{6}a^{2}+21\ell ^{4}a^{4}+7\ell
^{2}a^{6}+a^{8}\right) .  \label{EvacKerr9}
\end{equation}%
At once we notice a different value respect to the one found for the same
solution in Ref.\cite{Das}. The origin of this mismatch may be, in fact, that in order to obtain a quasilocal stress tensor, the authors of Ref.\cite%
{Das} performed an integration by parts in the highest-derivative terms of
the counterterm series in nine dimensions. This may lead to finite
contributions to the vacuum energy different from ours.

\section{\label{sec:Conclusions}Conclusions and prospects}

We have obtained explicit expressions for the vacuum energy for Kerr-AdS
black holes, geometry that admits a maximal number of $[(D-1)/2] $ commuting
axial symmetries. The expression up to nine dimensions exhibits an
interesting property: the zero-point energy reduces to the one of a static
AAdS black hole when all rotation parameters are taken as equal to each
other. It would be interesting to understand the implications in the
boundary CFT of this vacuum energy degeneracy. It is likely this fact
can be related to a symmetry-enhancement that the vacuum solution metric
should exhibit in that case (the deformation induced by the rotation parameters is the same
in all azimuthal directions).

We have not been able to identify a pattern in $E_{vac}$\ for Kerr-AdS,
which would allow us to pass from the particular results of the last section
to a general formula, valid in any odd dimension. We believe that the
explicit expressions we found can be cast in a more compact form using
parametric integrations.

On the other hand, the agreement between the results from Kounterterm
charges and the ones obtained by holographic techniques in AdS gravity
suggests that Eq.(\ref{q0ij}) should be a part of the stress tensor $%
T^{ij}[h]=(2/\sqrt{-h})\delta I_{ren}/\delta h_{ij}$, defined upon the
addition of local counterterms. A direct comparison between both formulas
would require, in general, converting extrinsic quantities into intrinsic
ones. This can be done considering the expansion of the extrinsic curvature
for AAdS spacetimes and noticing that all terms in Eq.(\ref{q0ij}) up to the
relevant order can be expressed as contractions between the Riemann and the
Schouten tensors of the boundary metric \cite{MOO}.

A non-zero value for $E_{vac}$ modifies the derivation of the positivity of
energy for asymtotically AdS spacetimes, as it has been emphasized in Ref.%
\cite{Cheng:2005wk}. The existence of globally defined Killing spinors in
a supersymmetry extension of AdS gravity results in a vacuum energy formula given in
terms of the coefficients of the Fefferman-Graham expansion of the metric \cite{fg}.
We hope that the ongoing efforts to compare the Cheng-Skenderis formula to ours
are able to provide an answer to this issue.

\begin{acknowledgments}
We are grateful to R. Aros, D. Astefanesei, O. Miskovic and P. Sloane for helpful remarks. This work was funded in part by FONDECYT Grants No.
1090357 and 1131075 (R.O.) and UNAB Grant DI-117-12/R. G.O. received support from CONICYT.
\end{acknowledgments}

\appendix

\section{Kronecker delta of rank $p$ \ \label{Delta}}

Many of the formulas in this paper are written in a more compact form thanks
to the use of the totally-antisymmetric Kronecker delta. Such an object of rank
$p$ is defined as the determinant
\begin{equation}
\delta _{\left[ \mu _{1}\cdots \mu _{p}\right] }^{\left[ \nu _{1}\cdots \nu
_{p}\right] }:=\left\vert
\begin{array}{cccc}
\delta _{\mu _{1}}^{\nu _{1}} & \delta _{\mu _{1}}^{\nu _{2}} & \cdots &
\delta _{\mu _{1}}^{\nu _{p}} \\
\delta _{\mu _{2}}^{\nu _{1}} & \delta _{\mu _{2}}^{\nu _{2}} &  & \delta
_{\mu _{2}}^{\nu _{p}} \\
\vdots &  & \ddots &  \\
\delta _{\mu _{p}}^{\nu _{1}} & \delta _{\mu _{p}}^{\nu _{2}} & \cdots &
\delta _{\mu _{p}}^{\nu _{p}}%
\end{array}%
\right\vert \,.
\end{equation}
A contraction of $k\leq p$ indices in the Kronecker delta of rank $p$
produces a delta of rank $p-k$,
\begin{equation}
\delta _{\left[ \mu _{1}\cdots \mu _{k}\cdots \mu _{p}\right] }^{\left[ \nu
_{1}\cdots \nu _{k}\cdots \nu _{p}\right] }\,\delta _{\nu _{1}}^{\mu
_{1}}\cdots \delta _{\nu _{k}}^{\mu _{k}}=\frac{\left( N-p+k\right) !}{%
\left( N-p\right) !}\,\delta _{\left[ \mu _{k+1}\cdots \mu _{p}\right] }^{%
\left[ \nu _{k+1}\cdots \nu _{p}\right] }\,,
\end{equation}%
where $N$ is the range of indices.

\section{Gauss-normal coordinate frame}

For most of the discussions in the present paper, the relevant components of
the Christoffel connection $\Gamma _{\mu \nu }^{\alpha }$ are expressed in
terms of the extrinsic curvature and radial derivatives of the lapse
function $N$ as
\begin{equation}
\Gamma _{ij}^{r}={\frac{1}{{N}}}K_{ij},\qquad \Gamma
_{rj}^{i}=-NK_{j}^{i},\qquad \Gamma _{rr}^{r}={\frac{N^{\prime }}{{N}}}.
\end{equation}

In our conventions, the Riemann tensor is defined as

\begin{equation}
R_{\mu \beta \nu }^{\alpha }=\partial _{\beta }\Gamma _{\nu \mu }^{\alpha
}-\partial _{\nu }\Gamma _{\beta \mu }^{\alpha }+\Gamma _{\beta \gamma
}^{\alpha }\Gamma _{\nu \mu }^{\gamma }-\Gamma _{\nu \gamma }^{\alpha
}\Gamma _{\beta \mu }^{\gamma }\,,
\end{equation}
what leads to the well-known Gauss-Codazzi relations
\begin{eqnarray}
&&R_{kl}^{jr}={\frac{1}{{N}}}\left( \nabla _{l}K_{k}^{i}-\nabla
_{k}K_{l}^{i}\right) , \\
&&R_{kr}^{ir}={\frac{1}{{N}}}\left( K_{k}^{i}\right) ^{\prime
}-K_{l}^{i}K_{k}^{l} , \\
&&R_{kl}^{ij}=\mathcal{R}_{kl}^{ij}(h)-K_{k}^{i}K_{l}^{j}+K_{l}^{i}K_{k}^{j} ,
\end{eqnarray}%
where $\nabla _{l}=\nabla _{l}(\Gamma )$ denotes the covariant derivative
defined in terms of the Christoffel \ symbol of the boundary $\Gamma
_{jk}^{i}=\Gamma _{jk}^{i}(h)$.

\section{Parametrization of the sphere $S^{2n-1}$}

\label{apC}

In $D=2n+1$ dimensions we have $n-1$ polar angles $\theta_{i}$, where $%
0\leq\theta_{i}\leq{\frac{\pi }{2}}$. Altogether, we have $n$ azimuthal
angles, $0\leq\phi_{i}\leq 2\pi$. The polar angles are related to the
direction cosines as

\begin{equation}
\mu_{i}=\prod_{j=1}^{i-1}\cos\theta_{j}\sin\theta_{i} ,
\end{equation}

or, more explicitly

\begin{equation}
\begin{array}{lcr}
\mu _{1}=\sin \theta _{1} &  &  \\
\mu _{2}=\cos \theta _{1}\sin \theta _{2} &  &  \\
\mu _{3}=\cos \theta _{1}\cos \theta _{2}\sin \theta _{3} &  &  \\
\vdots &  &  \\
\mu _{n-1}=\cos \theta _{1}\cos \theta _{2}\cdots \cos \theta _{n-2}\sin
\theta _{n-1} &  &  \\
\mu _{n}=\cos \theta _{1}\cos \theta _{2}\cdots \cos \theta _{n-2}\cos
\theta _{n-1}. &  &
\end{array}%
.
\end{equation}


\begin{thebibliography}{99}
\bibitem{E07} The Casimir energy for seven-dimensional rotating black hole was obtained in Ref.\cite{Gibbons2} using a coordinate
system for Kerr-AdS that leaves the asymptotic region static. As a consequence, the vacuum energy does not depend on
the rotation parameters and it coincides with the vale for the Schwarzschild black hole.
\end{thebibliography}
\end{document}